\documentclass{amsart}

\input{epsf}

\newtheorem{thm}{Theorem}[section]

\newtheorem{lem}[thm]{Lemma}

\theoremstyle{definition}

\theoremstyle{remark}

\numberwithin{equation}{section}

\begin{document}

\title[On the resolvent and spectral functions ...]
{On the resolvent and spectral functions of a second order
differential operator with a regular singularity}%
\author[H.\ Falomir, M.\ A.\ Muschietti and
P.\ A.\ G.\ Pisani]{H.\ Falomir$^A$, M.\ A.\ Muschietti$^B$ and
P.\ A.\ G.\ Pisani$^A$
}%

\address{$A)$ IFLP, Departamento de F{\'\i}sica - Facultad de Ciencias
Exactas, UNLP, C.C. 67 (1900) La Plata, Argentina  }

\address{$B)$ Departamento de Matem{\'a}tica - Facultad de Ciencias Exactas,
UNLP, C.C. 172 (1900) La Plata, Argentina}



\begin{abstract}
We consider the resolvent of a second order differential operator
with a regular singularity, admitting a family of self-adjoint
extensions. We find that the asymptotic expansion for the
resolvent in the general case presents unusual powers of $\lambda$
which depend on the singularity. The consequences for the pole
structure of the  $\zeta$-function, and for the small-$t$
asymptotic expansion of the heat-kernel, are also discussed.
\end{abstract}

\maketitle
\section{Introduction}

It is well known that in Quantum Field Theory under external
conditions, quantities like vacuum energies and effective actions,
which describe the influence of boundaries or external fields on
the physical system, are generically divergent and require a
renormalization to get a physical meaning.

In this context, a powerful and elegant regularization scheme to
deal with these problems is based on the use of the
$\zeta$-function \cite{Dowker,Hawking} or the heat-kernel (for
recent reviews see, for example,
\cite{Elizalde,Bytsenko,Klaus,Bordag,Dmitri}) associated to the
relevant differential operators appearing in the quadratic part of
the actions. In this way, ground state energies, heat-kernel
coefficients, functional determinants and partition functions for
quantum fields can be given in terms of the corresponding
$\zeta$-function, where the ultraviolet divergent pieces of the
one-loop contributions are encoded as poles of its holomorphic
extension.

Thus, it is of major interest in Physics to determine the
singularity structure of $\zeta$-functions associated with these
physical models.

\bigskip

In particular \cite{Seeley}, for an elliptic boundary value
problem in a $\nu$-dimensional compact manifold with boundary,
described by a differential operator $A$ of order $\omega$, with
smooth coefficients and a ray of minimal growth, defined on a
domain of functions subject to local boundary conditions, the
$\zeta$-function
\begin{equation}\label{zeta-func-def}
  \zeta_A(s):= Tr\{A^{-s}\}
\end{equation}
has a meromorphic extension to the complex $s$-plane whose
singularities are isolated simple poles at $s=(\nu-j)/\omega$,
with $j=0,1,2,\dots$

In the case of positive definite operators, the $\zeta$-function
is related, via Mellin transform, to the trace of the heat-kernel
of the problem,  and the pole structure of $\zeta_A(s)$ determines
the small-$t$ asymptotic expansion of this trace
\cite{Seeley,Gilkey}:
\begin{equation}\label{heat-trace}
  Tr\{e^{-t A}\}\sim \sum_{j=0}^\infty a_j(A)\, t^{(j-\nu)/\omega},
\end{equation}
where the coefficients are related to the residues by
\begin{equation}\label{coef-res}
  a_j(A)=\left.{\rm Res}\right|_{s=(\nu-j)/\omega}
  \Gamma(s)\,\zeta_A(s).
\end{equation}

\bigskip

For operators of the form $-\partial_x^2 + V(x)$ with a singular
potential $V(x)$ asymptotic to $\kappa/ x^{2}$ as $x\rightarrow
0$, this expansion is substantially different. If $\kappa \geq
3/4$, the operator is essentially self-adjoint. This case has been
treated in \cite{Callias1,Callias2,Callias3}, where log terms are
found, as well as terms with coefficients which are distributions
concentrated at the singular point $x=0$. For the case $\kappa >
-1/4$, the Friedrichs extension has been treated in
\cite{Bruening} for operators in $\mathbf{L_2}(0,1)$, and in
\cite{Bruening-Seeley} for operators in
$\mathbf{L_2}(\mathbf{R^+})$, making use of the scale invariance
of the operator domain and explicit representations of the
resolvent. Moreover, as a particular case of a manifold with an
isolated conic singularity, reference \cite{Mooers} gave a
description of the boundary behavior of the Friedrichs heat-kernel
which does not make use of the resolvent, and showed v{\'\i}a boundary
maps how it can be used to construct the heat-kernel for other
self-adjoint extensions of these operators, showing explicitly the
first two terms in the asymptotic expansion of the trace of their
difference.

\bigskip

On the other hand, reference \cite{FPW} gave the pole structure of
the $\zeta$-function of a second order differential operator
defined on the (non compact) half-line $\mathbf{R}^+$,  having a
singular zero-th order term $V(x)=\kappa \, x^{-2}+ x^2$. It
showed that, for a certain range of real values of $\kappa$, this
operator admits nontrivial self-adjoint extensions in
$\mathbf{L_2}(\mathbf{R^+})$, for which the associated
$\zeta$-function (given by an integral representation) presents
isolated simple poles which (in general) do not lie at $s=(1-j)/2$
for $j=0,1,\dots$ (as would be the case for a regular $V(x)$), and
can even take irrational values.

A similar structure has been noticed in \cite{FMPS} for the
singularities of the $\zeta$ and $\eta$-functions of a system of
first order differential operators with a singular zero-th order
term $\sim g\, x^{-1}$, which also admits a family of self-adjoint
extensions for real $g$ taking  values in certain range. It has
been shown that, in the general case, the asymptotic expansion of
the resolvent contains $g-$dependent powers of $\lambda$ which
make the $\zeta$ and $\eta$-functions to present poles lying at
points which depend on the singularity, with residues depending on
the self-adjoint extension.

\bigskip

Let us mention that singular potentials $\sim 1/x^2$ have been
considered in the description of several physical systems, like
the Calogero Model \cite{Calogero,Perelomov,FPW,Basu}, conformal
invariant quantum mechanical models \cite{DA-F-F,Camblong,Coon}
and, more recently, the dynamics of quantum particles in the
asymptotic  near-horizon region of black-holes
\cite{Claus,Gibbons,Govindarajan,Birmingham,Moretti}. The
self-adjoint extensions of these operators have also been
considered in \cite{Tsutsui}. Moreover, singular superpotentials
has been considered as possible agents of supersymmetry breaking
in models of Supersymmetric Quantum Mechanics
\cite{Jevicki,Pernice,Das}.

\bigskip

It is the aim of the present article to analyze the behavior of
the resolvent, the $\zeta$-function and the trace of the
heat-kernel of a second order differential operator with a regular
singularity in a compact segment, $D_x=-\partial_x^2+{g(g-1)}\,
{x^{-2}}$, for those values of $g$ for which it admits a family of
self-adjoint extensions.

Following the scheme developed in \cite{FMPS}, we will show that
the asymptotic expansion for the resolvent in the general case
presents powers of $\lambda$ which depend on the singularity, and
can even take irrational values. The consequence of this behavior
on the corresponding $\zeta$-function is the presence of simple
poles lying at points which also depend on the singularity, with
residues depending on the self-adjoint extension considered.

{We first construct the resolvents for two particular extensions,
for which the boundary condition at the singular point $x=0$ is
invariant under the scaling $x \rightarrow c\,x$. The resolvent
expansion for these special extensions displays the usual powers,
leading to the usual poles for the $\zeta$-function (and the usual
structure for the asymptotic expansion of the heat-kernel trace).

The resolvents of the remaining extensions are convex linear
combinations of these special extensions, but the coefficients in
the convex combination depend on the eigenvalue parameter
$\lambda$. This dependence leads to unusual powers in the
resolvent expansion, and hence to unusual poles for the
zeta-function (and unusual powers in the asymptotic expansion of
the heat-kernel trace).

These self-adjoint extensions are not invariant under the scaling
$x\rightarrow c\,x$. As $c \rightarrow 0$ they tend (at least
formally) to one of the invariant extensions, and as $c
\rightarrow \infty$ they tend to the other. As $c \rightarrow 0$
the residues at the anomalous poles tend to zero, whereas as $c
\rightarrow \infty$ these residues become infinite. The way these
residues depend on the boundary condition is explained by a
scaling argument in Section 7.}

\bigskip

The structure of the article is as follows: In Section
\ref{the-operator} we define the operator and determine its
self-adjoint extensions for $\frac 1 2 < g < \frac 3 2$, and in
Section \ref{the-spectrum} we study their spectra. In Section
\ref{the-resolvent} we construct the resolvent for a general
extension as a linear combination of the resolvent of two limiting
cases, and in Section \ref{trace-resolvent} we consider the traces
of these operators. The asymptotic expansions of these traces,
evaluated in Section \ref{Asymptotic-expansion}, are used in
Section \ref{spectral-functions} to construct the associated
$\zeta$-function and study its singularities, as well as the
small-t asymptotic expansion of the heat-kernel trace. The special
case $g=\frac 1 2$ is considered in Appendix \ref{g=1/2}.

\section{The operator and its self-adjoint extensions}
\label{the-operator}

Let us consider the differential operator
\begin{equation}\label{D}
  D_x=-\frac{d^2}{dx^2}+\frac{g(g-1)}{x^2}\, ,
\end{equation}
with $g\in \mathbb{R}$, defined on a domain of smooth functions
with compact support in a segment, $\mathcal{D}(D)=
\mathcal{C}_0^\infty(0,1)$. It can be easily seen that $D_x$ so
defined is symmetric.

The adjoint operator $D_x^*$, which is the maximal extension of
$D_x$, is defined on the domain $\mathcal{D}(D_x^*)$ of functions
$\phi(x)\in \mathbf{L_2}(0,1)$, having a locally sumable second
derivative and such that
\begin{equation}\label{DPhi}
    D_x\phi(x)=-\phi''(x)+ \frac{g(g-1)}{x^2}\,
    \phi(x)=f(x)\in \mathbf{L_2}(0,1)\, .
\end{equation}

\bigskip

\begin{lem} \label{lema1-1}
If $\phi(x)\in \mathcal{D}(D_x^*)$ and $\frac{1}{2}< g<
\frac{3}{2}$, then\footnote{The case $g= \frac 1 2$ will be
considered separately, in Appendix \ref{g=1/2}.}
\begin{equation}\label{lemaI-1}
    \displaystyle{\left|\,\phi(x)-\left(\frac{C_1[\phi]\, x^g +
  C_2[\phi]\, x^{1-g}}{\sqrt{2g-1}}\right)\right|
  \leq \frac{\|D_x\phi(x)\|}{(3/2 -g)\sqrt{2g+1}} \  x^{3/2}}
\end{equation}
and
\begin{equation}\label{lemaI-2}
    \displaystyle{
      \left|\,\phi'(x)-\left(\frac{g\, C_1[\phi]\, x^{g-1} +
  (1-g)\, C_2[\phi]\, x^{-g}}{\sqrt{2g-1}}\right)\right| \leq
  \frac{3/2 \, \|D_x\phi(x)\|}{(3/2 -g)\sqrt{2g+1}} \  x^{1/2}}
\end{equation}
for some constants $C_1[\phi]$ and $C_2[\phi]$, where $\| \cdot
\|$ is the $\mathbf{L_2}$-norm.
\end{lem}

\noindent {\bf Proof:} Let us write $\phi(x)= x^g u(x)$. Then,
Eq.\ (\ref{DPhi}) implies
\begin{equation}\label{phi-chi-en0}
    \begin{array}{c} \displaystyle{
      u'(x)= K_2 \, x^{-2g} - x^{-2g} \, \int_0^x y^{g}\, f(y)\, dy
     \,  ,} \\ \\ \displaystyle{
     u(x)=K_1+ \frac{K_2}{1-2g} \, x^{1-2g} -
      \int_0^x y^{-2g} \, \int_0^y z^{g}\, f(z)\, dz \ dy
     \, ,}
    \end{array}
\end{equation}
for some constants $K_1$ and $K_2$. Now, taking into account that
\begin{equation}\label{schwarz}
    \begin{array}{c}\displaystyle{
      \left|\int_0^x y^{g} \, f(y)\, dy\right|\leq
      \frac{x^{g+1/2}}{\sqrt{2g+1}}\, \|f\| }\, ,
      \\ \\ \displaystyle{
      \left|\int_0^x y^{-2g} \int_0^y z^g\, f(z)\, dz\, dy\right|\leq
      \frac{x^{3/2-g}}{(3/2 -g)\sqrt{2g+1}}\, \|f\| }\,  ,
    \end{array}
\end{equation}
we immediately get Eqs.\ (\ref{lemaI-1}) and (\ref{lemaI-2}).

\bigskip

\begin{lem}
Let $\phi(x),\psi(x)\in \mathcal{D}(D^*)$ and $\frac{1}{2}< g<
\frac{3}{2}$. Then
\begin{equation}\label{DDstar}\begin{array}{c}
    \left(D_x \psi, \phi\right) -
  \left(\psi, D_x \phi\right) = \\ \\
  = \Big\{
  C_1[\psi]^* C_2[\phi] - C_2[\psi]^* C_1[\phi]\Big\}+
  \Big\{\psi(1)^*\, \phi'(1)
  - \psi'(1)^*\, \phi(1) \Big\}\, .
\end{array}
\end{equation}

\end{lem}

\noindent {\bf Proof:} From Eq.\ (\ref{DPhi}) one easily obtains
\begin{equation}\label{DDstar2}\begin{array}{c}
      \left(D_x \psi, \phi\right) -
  \left(\psi, D_x \phi\right) = \\ \\
    =\displaystyle{
    \lim_{\varepsilon \rightarrow 0^+}\int_\varepsilon^1
    \partial_x \Big\{\psi(x)^*\, \phi'(x)
    - \psi'(x)^* \,\phi(x)
    \Big\} \, dx}\, ,
\end{array}
\end{equation}
from which, taking into account the results in Lemma
\ref{lema1-1}, Eq.\ (\ref{DDstar}) follows directly.

\bigskip

Now, if $\psi(x)$ in Eq.\ (\ref{DDstar}) belongs to the domain of
the closure of $D_x$, $\overline{D}_x=(D_x^{*})^*$,
\begin{equation}\label{en-la-clausura}
  \psi(x)\in
\mathcal{D}(\overline{D}_x) \subset \mathcal{D}(D_x^*)\,  ,
\end{equation}
then the right hand side of Eq.\ (\ref{DDstar}) must vanish for
any $\phi(x)\in \mathcal{D}(D_x^*)$. Therefore
\begin{equation}\label{Psi-clausura}
    C_1[\psi]=C_2[\psi]=\psi(1)=\psi'(1)=0\, .
\end{equation}

\bigskip

On the other hand, if $\psi(x),\phi(x)$ belong to the domain of a
symmetric extension of $D_x$ (contained in $\mathcal{D}(D_x^*)$),
the right hand side of Eq.\ (\ref{DDstar}) must also vanish.

Thus, the closed extensions of $D_x$ correspond to the subspaces
of $\mathbb{C}^4$ under the map $\Phi\rightarrow \left( C_1[\Phi],
C_2[\Phi], \phi(1), \phi'(1) \right)$, and the self-adjoint
extensions correspond to those subspaces $S\subset \mathbb{C}^4$
such that $S=S^{\perp}$, with the orthogonal complement taken in
the sense of the symplectic  form on the right hand side of Eq.\
(\ref{DDstar}).

\bigskip

For definiteness, in the following we will consider self-adjoint
extensions satisfying the local boundary condition
\begin{equation}\label{BC1}
  \phi(1)=0\, .
\end{equation}
Each such extension is determined by a condition of the form
\begin{equation}\label{BC2}
    \alpha\, C_1[\Phi] + \beta\, C_2[\Phi] = 0\, ,
\end{equation}
with $\alpha,\beta \in \mathbb{R}$, and $\alpha^2 + \beta^2 =1$.
We denote this extension by $D_x^{(\alpha, \beta)}$.

\section{The spectrum} \label{the-spectrum}

In order to determine the spectrum of the self-adjoint extensions
of $D_x$ for $\frac{1}{2}< g< \frac{3}{2}$, we need the solutions
of
\begin{equation}\label{Ec-hom}
  (D_x-\lambda)\phi_\lambda(x)=0 \, ,
\end{equation}
satisfying the boundary conditions in Eqs.\ (\ref{BC1}) and
(\ref{BC2}).

\bigskip

The general solution of the homogeneous equation for $\lambda=0$
is
\begin{equation}\label{lambda0}
  \phi_0(x)=\frac{1}{\sqrt{2g-1}} \left(
  C_1\, x^g +
  C_2\, x^{1-g}\right)\, ,
\end{equation}
and the boundary conditions in Eqs.\ (\ref{BC1}) and (\ref{BC2})
imply that
\begin{equation}\label{BBBCCC}
  C_1+C_2=0\, , \quad \alpha\,C_1 + \beta\,C_2=0 \, .
\end{equation}
Consequently, there are no zero modes except for the self-adjoint
extension characterized by $\alpha =\beta=1/\sqrt{2}$.

\bigskip

For $\lambda\neq 0$, the solutions of Eq.\ (\ref{Ec-hom}) are of
the form
\begin{equation}\label{sol-hom-1}\begin{array}{c}
  \displaystyle{  \phi(x) =
  \frac{{C}_1}{{\sqrt{2\,g-1}}}\,
    \frac{
       \Gamma(\frac{1}{2} + g)}{
       2^{ \frac{1}{2}-g}\,{\mu }^
        {g-\frac{1}{2} }}\,{\sqrt{x}}\,
       J_{g-\frac{1}{2}}(\mu\,x )
       +}
  \\ \\
  \displaystyle{+\frac{{C}_2}{{\sqrt{2\,g-1}}}\,\frac{
       \Gamma(\frac{3}{2} - g)}{
       2^{g-\frac{1}{2}}\,{\mu }^
        { \frac{1}{2} - {g} }}\,{\sqrt{x}} \,
       J_{\frac{1}{2} - g}(\mu\,x )\, ,}
\end{array}
\end{equation}
where $\mu = +\sqrt{\lambda}$, and the $\mu$-dependent
coefficients have been introduced for later convenience.

Taking into account that
\begin{equation}\label{J-Bessel}
  J_\nu(z)= z^{\nu }\,\left\{ \frac{1}
     {2^{\nu }\,
       \Gamma(1 + \nu )}
     + O(z^2) \right\}\, ,
\end{equation}
we get from Eqs.\ (\ref{lemaI-1}) and (\ref{BC2})
\begin{equation}\label{ec-spectrum}
    \alpha\, C_1 + \beta\, C_2=0\, .
\end{equation}

On the other hand, the condition in Eq.\ (\ref{BC1}) implies
\begin{equation}\label{sol-hom-1-en1}\begin{array}{c}
  \displaystyle{  \phi(1) =
  \frac{{C}_1}{{\sqrt{2\,g-1}}}\,
    \frac{
       \Gamma(\frac{1}{2} + g)}{
       2^{ \frac{1}{2}-g}\,{\mu }^
        {g-\frac{1}{2} }}\,
       J_{g-\frac{1}{2}}(\mu)
       +}
  \\ \\
  \displaystyle{+\frac{{C}_2}{{\sqrt{2\,g-1}}}\,\frac{
       \Gamma(\frac{3}{2} - g)}{
       2^{g-\frac{1}{2}}\,{\mu }^
        { \frac{1}{2} - {g} }}\,
       J_{\frac{1}{2} - g}(\mu) =0\, .}
\end{array}
\end{equation}

\bigskip

For $\alpha = 0$, Eq.\ (\ref{ec-spectrum}) implies $C_2 = 0$
(Dirichlet boundary conditions at the origin). Therefore,
$\phi(1)=0 \Rightarrow  J_{g-\frac 1 2}(\mu)=0$. Thus, the
spectrum of this self-adjoint extension is positive and
non-degenerate, with the eigenvalues of $D_x^D:=D_x^{(0,1)}$ given
by
\begin{equation}\label{alpha0}
  \lambda_{n} = j_{g-\frac 1 2,n}^2\, ,
   \quad n=1,2,\dots\, ,
\end{equation}
where $j_{\nu,n}$ is the $n$-th positive zero of the Bessel
function $J_{\nu}(z)$\footnote{\label{zeroes} Let us recall that
large zeros of $J_{\nu}(\lambda)$ have the asymptotic expansion
\begin{equation}\label{large-zeroes}
  j_{\nu,n}\simeq \gamma -\frac{4 \nu^2-1}{8 \gamma}+O\left(\frac 1
  \gamma\right)^3\, ,
\end{equation}
with $\gamma=\left( n+\frac{\nu}{2}-\frac{1}{4} \right) \pi$.}.

\bigskip

For $\alpha\neq 0$, from Eqs.\ (\ref{ec-spectrum}) and
(\ref{sol-hom-1-en1}) we easily get the following transcendental
equation for the eigenvalues of $D_x^{(\alpha,\beta)}$:
\begin{equation}\label{spectrum}
    F(\mu):={\mu }^{2\,g-1}\,\frac{
    J_{\frac{1}{2} - g}(\mu )}
    {J_{g-\frac{1}{2}}(\mu )}
   = \rho(\alpha,\beta)
        \, ,
\end{equation}
where we have defined
\begin{equation}\label{rho}
  \rho(\alpha,\beta):= \frac{\beta }{\alpha}\  \frac{2^{ 2\,g-1}\,
    \Gamma(\frac{1}{2} + g)}{
    \Gamma(\frac{3}{2} - g)} \, .
\end{equation}

\begin{figure}
    \epsffile{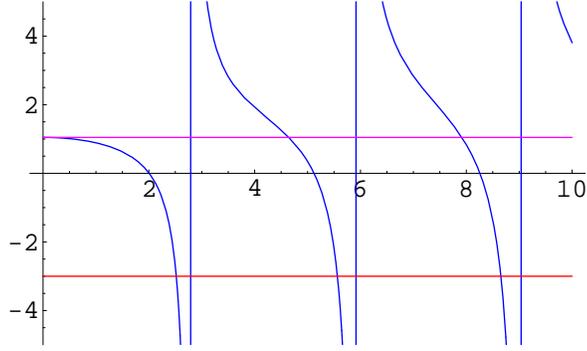}
    \caption{Plot for $F(\mu)$, $\rho(\alpha,\beta)=-3$ and
    $\displaystyle{\rho(\alpha,\alpha)}$, with $g=3/4$.}
    \label{figure}
\end{figure}

For the positive eigenvalues ${\lambda}=\mu^2$, both sides in Eq.\
(\ref{spectrum})  have been plotted in Figure \ref{figure}, for
particular values of $\rho(\alpha,\beta)$ and $g$.

\begin{figure}
    \epsffile{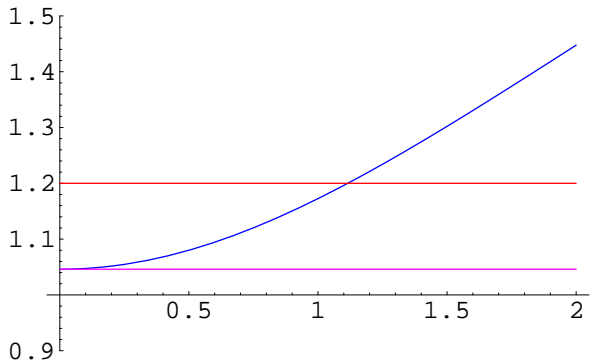}
    \caption{Plot for $F(i\,\mu)$, $\rho(\alpha,\beta)=1.2$ and
    $\displaystyle{\rho(\alpha,\alpha)}$, with $g=3/4$.}
    \label{lambda-neg}
\end{figure}

Moreover, if $\beta/\alpha >1 \Rightarrow
\rho(\alpha,\beta)>\rho(\alpha,\alpha)$, and the extension
$D_x^{(\alpha,\beta)}$ has a negative eigenvalue. Indeed, if
$\lambda_- = (i \mu)^2 <0$, then
\begin{equation}\label{Fdeimu}\begin{array}{c}
  \displaystyle{F(i\,\mu)={\mu }^{ 2\,g-1}\,
  \frac{
    I_{\frac{1 }{2}- g}(\mu )}
    {I_{g-\frac{1}{2}}(\mu )}=} \\ \\
  \displaystyle{ =2^{2\,g-1}\,
    \frac{
     \Gamma(\frac{1}{2} + g)}{\Gamma(\frac{3}{2} - g)}
     \left\{
     1 + \frac{\left( 2\,g-1   \right) \,{\mu }^2}
   {\left( 3 - 2\,g \right) \,
     \left( 1 + 2\,g \right) }
   +{O}(\mu^4 )\right\} }\, ,
\end{array}
\end{equation}
where $I_\nu(\mu)$ is the modified Bessel function. For a plot,
see Figure \ref{lambda-neg}\footnote{It can be seen that this
negative eigenvalue goes to $-\infty$ as $\alpha \rightarrow 0$,
while  the corresponding eigenfunction tends to concentrate on the
singularity at $x=0$. See also \cite{Asorey}.}.

\bigskip

Notice that the spectrum is always non-degenerate, and there is a
positive eigenvalue between each pair of consecutive squared
zeroes of $J_{g-\frac 1 2}(\lambda)$. Therefore, from Eq.\
(\ref{large-zeroes}) we get $\lambda_n = \pi^2 \, n^2 + O(n)$.

\bigskip

In particular, for the $\beta=0$ extension (which we call the
``N-extension"), $D_x^N:=D_x^{(1,0)}$, it can be seen fron Eq.\
(\ref{spectrum}) that the eigenvalues are given by
\begin{equation}\label{eigen-beta0}
  \lambda_{n} =  j_{\frac 1 2 -g ,n}^2\, , \ n=1,2,\dots\, ,
\end{equation}
where $j_{\frac 1 2 -g ,n}^2$ are the positive zeroes of
$J_{\frac{1}{2} - g}(\mu )$.

\section{The resolvent} \label{the-resolvent}

In this Section we will construct the resolvent of $D_x$,
\begin{equation}\label{def-resolv}
  G(\lambda)=(D_x-\lambda)^{-1}\, ,
\end{equation}
for its different self-adjoint extensions when $\frac{1}{2}< g<
\frac{3}{2}$.

We will first consider the two limiting cases in Eq.\ (\ref{BC2}),
namely the ``$D$-extension", for which $\alpha=0 \Rightarrow
C_2[\phi] =0$, and the ``$N$-extension", with $\beta=0 \Rightarrow
C_1[\phi]=0$. The resolvent for a general self-adjoint extension
will be later evaluated as a linear combination of those obtained
for these two limiting cases.

\bigskip

For the kernel of the resolvent we have
\begin{equation}\label{G}
   (D_x-\mu^2)\, G(x,y; \mu^2) =
\delta(x-y)\, ,
\end{equation}
where $\mu^2=\lambda $, with $-\pi/ 2 < {\rm arg} (\mu) \leq \pi/
2$.

To proceed, we need some particular solutions of the homogeneous
equation (\ref{Ec-hom}). Then, let us define
\begin{equation}\label{soluciones}
    \left\{
  \begin{array}{l}
    \displaystyle{L^D(x,\mu)= {\sqrt{x}}\,
       J_{g-\frac{1}{2}}(\mu\,x )\,  ,}\\ \\
    \displaystyle{L^N(x,\mu)=
    {\sqrt{x}}\,
       J_{\frac{1}{2}-g}(\mu\,x )\,  ,}\\ \\
    \displaystyle{R(x,\mu)=
    {\sqrt{x}}\,\left( J_{\frac{1}{2} - g}(\mu)\,
     J_{g-\frac{1}{2}}(\mu\,x )
     - J_{g-\frac{1}{2}}(\mu)
     \,J_{\frac{1}{2} - g}(\mu \,x )
     \right)
    \, .}
        \end{array}
    \right.
\end{equation}
Notice that $R(1,\mu)=0$.

We will also need the Wronskians
\begin{equation}\label{W-D}
\left\{
\begin{array}{c}
  \displaystyle{
  W\left[ L^D(x,\mu), R(x,\mu) \right] =
  \frac{2\,\cos (g\,\pi )}
    {\pi } \,J_{g-\frac{1}{2}}(\mu )
   =\frac{1}{ \gamma_D(\mu)}\, ,
    } \\ \\
   \displaystyle{
  W\left[ L^N(x,\mu), R(x,\mu) \right] =
  \frac{2\,\cos (g\,\pi )}{\pi }\,J_{\frac{1}{2} - g}(\mu )
    =\frac{1}{ \gamma_N(\mu)}\, ,
    }
    \end{array}
  \right.
\end{equation}
which  vanish only at the zeroes of $J_{\nu}({\mu})$, for $\nu =
\pm\left(g-\frac 1 2\right)$.

\bigskip

\subsection{The resolvent for the $D$-extension} In this case,
the function
\begin{equation}\label{func-inhom}
  \phi(x) = \int_0^1 G_D(x,y; \mu^2)\, f(y)\, dy
\end{equation}
must satisfy $\phi(1)=0$ and $C_2[\phi]=0$, for any function
$f(x)\in \mathbf{L_2}(0,1)$.

This requires that
\begin{equation}\label{GD11}
  G_D (x,y;\mu^2)= \gamma_D(\mu) \times  \left\{
  \begin{array}{c}
    L^D(x,\mu)\, R(y, \mu),\ {\rm for}\ x\leq y\, , \\ \\
    R(x,\mu)\, L^D(y,\mu), \ {\rm for}\ x \geq y\, .
  \end{array}\right.
\end{equation}
The fact that the boundary conditions are satisfied, as well as
$(D_x-\mu^2)\, \phi(x)=f(x)$, can be straightforwardly verified
from Eqs.\ (\ref{soluciones}) and (\ref{W-D}).

\bigskip

Indeed, from Eqs.\ (\ref{func-inhom}),  (\ref{GD11}),
(\ref{soluciones}) and (\ref{W-D}), one gets
\begin{equation}\label{near0-D}
    \phi(x)=
    \frac{C_1^D[\phi]}{\sqrt{2\,g-1}} \, x^g + O({x}^{3/2})\, ,
\end{equation}
with
\begin{equation}\label{C+}
  C_1^D[\phi] =   \frac{\pi\, \mu^{g-\frac 1 2
     }\, \sqrt{2\,g-1} }
    {2^{\frac{1}{2}+ g} \cos(g\, \pi)
    J_{g-\frac{1}{2}}({\mu}) \,
    \Gamma\left(\frac{1}{2} +g\right)}
    \ \int_0^1 R(y, \mu) f(y) \, dy \, ,
\end{equation}
for $\mu$ not a zero of $J_{g-\frac{1}{2}}({\mu})$.


Notice that $C_1^D[\phi]\neq 0$ if the integral in the right hand
side of Eq.\ (\ref{C+}) is non vanishing.

\bigskip

\subsection{The resolvent for the $N$-extension} In this case,
the function
\begin{equation}\label{func-inhom-N}
  \phi(x) = \int_0^1 G_N( x,  y; \mu^2)\, f(y)\, dy
\end{equation}
must satisfy $\phi(1)=0$ and $C_1[\phi]=0$, for any function
$f(x)\in \mathbf{L_2}(0,1)$.

This requires that
\begin{equation}\label{GN11}
  G_N (x,y;\mu^2)= \gamma_N(\mu) \times  \left\{
  \begin{array}{c}
    L^N(x,\mu)\, R(y,\mu),\ {\rm for}\ x\leq y\, , \\ \\
    R(x,\mu)\, L^N(y,\mu), \ {\rm for}\ x \geq y\, .
  \end{array}\right.
\end{equation}
These boundary conditions, as well as the fact that $(D_x-\mu^2)\,
\phi(x)=f(x)$, can be straightforwardly verified from Eqs.\
(\ref{soluciones}) and (\ref{W-D}).

\bigskip

In this case, from Eqs.\ (\ref{func-inhom-N}), (\ref{GN11}),
(\ref{soluciones}) and (\ref{W-D}), one gets
\begin{equation}\label{near0-N}
    \phi(x)= \frac{C_2^N[\phi]}{\sqrt{2\,g-1}}
    \, x^{1-g} + O({x}^{3/2})\, ,
\end{equation}
with
\begin{equation}\label{C-}\begin{array}{c}
  \displaystyle{C_2^N[\phi] =   \frac{\pi\, \mu^{\frac 1 2 -g}
  \, {\sqrt{2\,g-1}}}
    {2^{\frac{3}{2}- g} \cos(g\, \pi)
    J_{\frac{1}{2}-g}({\mu}) \,
    \Gamma\left(\frac{3}{2}-g\right)}
    \int_0^1 R(y,\mu) f(y)
    \, dy\, ,}
\end{array}
\end{equation}
for $\mu$ not a zero of $J_{\frac{1}{2}-g}({\mu})$.


Notice that $C_2^N[\phi]\neq 0$ if the integral in the right hand
side of Eq.\ (\ref{C-}) (the same integral as the one appearing in
the $D$-extension, Eq.\ (\ref{C+})) is non vanishing.

\bigskip

\subsection{The resolvent for a general self-adjoint extension of $D_x$}

For the general case, we can adjust the boundary conditions
\begin{equation}\label{BC-general}
  \phi(1)=0\, ,\quad \alpha\, C_1[\phi] + \beta\, C_2[\phi] =
  0\, ,\ \alpha,\beta\neq 0\, ,
\end{equation}
for
\begin{equation}\label{func-inhom-gen}
  \phi(x) = \int_0^1 G( x,  y;\lambda)\, f(y)\, dy\, ,
\end{equation}
for any $f(x)\in \mathbf{L_2}(0,1)$, by taking a linear
combination of the resolvent for the limiting cases,
\begin{equation}\label{linear-comb}
  G(x,y; \lambda)= \left[1- \tau(\lambda)\right] G_D(x,y; \lambda) +
  \tau(\lambda)\, G_N(x,y;\lambda)\, .
\end{equation}

Since the boundary condition at $x=1$ is automatically fulfilled,
one must just impose
\begin{equation}\label{ec-tau}
  \alpha \left[1- \tau(\lambda)\right] C_1^D[\phi]
  + \beta\, \tau(\lambda)\, C_2^N[\phi] =0\, .
\end{equation}

Notice that, in view of Eq.\ (\ref{C+}), (\ref{C-}) and
(\ref{spectrum}),
\begin{equation}\label{nocero}
  \alpha \, C_1^D[\phi]-\beta\,  C_2^N[\phi]=0
\end{equation}
precisely when $\lambda=\mu^2$ is an eigenvalue of
$D_x^{(\alpha,\beta)}$. Therefore, from Eq.\ (\ref{ec-tau}) we get
the resolvent of $D_x^{(\alpha,\beta)}$ by setting
\begin{equation}\label{taudelambda}\begin{array}{c}
    \tau(\mu^2) = \displaystyle{\frac{\alpha \, C_1^D[\phi]}
  {\alpha \, C_1^D[\phi]-\beta\,  C_2^N[\phi]}} =
  \\  \\ =
  \displaystyle{
  \frac{1}{1 - \rho (\alpha ,\beta )\,{\mu }^{1 - 2\,g}\,
  \displaystyle{
  \frac{
       J_{g-\frac{1}{2}}(\mu )}
       {J_{\frac{1 }{2}-g}(\mu )}} }   }\, ,
\end{array}
\end{equation}
for  $\mu$ not a zero of $J_{\frac{1}{2}-g}(\mu )$.

\bigskip

\section{The trace of the resolvent} \label{trace-resolvent}

It follows from Eq.\ (\ref{linear-comb}) that the resolvent of a
general self-adjoint extension of $D_x$ can be expressed in terms
of the resolvents of the two limiting cases, $G_D(\lambda)$ and
$G_N(\lambda)$. Moreover, since the eigenvalues of any extension
grow as $n^2$ (see Section \ref{the-spectrum}), these resolvents
are trace class operators.

Then, we have
\begin{equation}\label{trazaG}
  Tr\left\{G(\lambda)\right\}=
  Tr\left\{G_D(\lambda)\right\}
   - \tau(\lambda)\left[
   Tr\left\{G_D(\lambda)\right\}-
   Tr\left\{G_N(\lambda)\right\}\right]
\end{equation}

\bigskip

From Eqs.\ (\ref{GD11}) and (\ref{GN11}) we straightforwardly get
(see Appendix \ref{integrals} for the details)
\begin{equation}\label{traza-GD}\begin{array}{c}
        Tr\{ G_D(\mu^2)\} = \displaystyle{
      \int_0^1
      tr\{
      G_D(x,x; \mu^2)\}\, dx} =
    \\ \\
    = \displaystyle{\frac{J_{\frac{1}{2} + g}(\mu )}
  {2\,\mu \,J_{g-\frac{1}{2}}(\mu )}
  =\frac{2\,g-1}{4\,\mu^2}-
  \frac{J'_{ g-\frac{1}{2} }(\mu )}
  {2\,\mu \,J_{g-\frac{1}{2}}(\mu )} \, ,}
\end{array}
\end{equation}
and
\begin{equation}\label{traza-GN}\begin{array}{c}
        Tr\{ G_N(\mu^2)\} = \displaystyle{
      \int_0^1
      tr\{
      G_N(x,x; \mu^2)\}\, dx} =
    \\ \\
    = \displaystyle{\frac{J_{\frac{3}{2} - g}(\mu )}
  {2\,\mu \,J_{\frac{1}{2} - g}(\mu )}
  =-\frac{2\,g-1}{4\,\mu^2}-
  \frac{J'_{\frac{1}{2} - g}(\mu )}
  {2\,\mu \,J_{\frac{1}{2} - g}(\mu )} \, ,}
\end{array}
\end{equation}
where we have taken into account that
\begin{equation}\label{numasmenos1}
  J_{\nu+ 1}(z) = \frac \nu z \, J_\nu(z)
  - J'_\nu(z)\, .
\end{equation}

Finally, we get
\begin{equation}\label{TrG2}\begin{array}{c}
  Tr\{G(\mu^2)\} =\displaystyle{
  \left\{\frac{2\,g-1}{4\,\mu^2}-
  \frac{J'_{ g-\frac{1}{2} }(\mu )}
  {2\,\mu \,J_{g-\frac{1}{2}}(\mu )}\right\} -}
  \\ \\
  \displaystyle{ - \tau(\mu^2) \left\{
  \frac{2\,g-1}{2\,\mu^2}-\frac{1}{2\,\mu}
  \left(
  \frac{J'_{ g-\frac{1}{2} }(\mu )}
  {J_{g-\frac{1}{2}}(\mu )}
  - \frac{J'_{\frac{1}{2} - g}(\mu )}
  {J_{\frac{1}{2} - g}(\mu )}
  \right) \right\} }\, .
\end{array}
\end{equation}

\bigskip

\section{Asymptotic expansion for the trace of the
resolvent}\label{Asymptotic-expansion}

Using the Hankel asymptotic expansion for Bessel functions
\cite{A-S} (see Appendix \ref{Hankel}), we get for the first term
in the right hand side of Eq.\ (\ref{TrG2})
\begin{equation}\label{asymp-trGD-upp}\begin{array}{c}
    \displaystyle{Tr\{ G_D(\mu^2)\} \sim
  \sum_{k=1}^\infty \frac{A_k(g,\sigma)}{\mu^k}
  = }\\ \\ =\displaystyle{
  \frac{i \,\sigma }{2\,\mu } +
  \frac{g}{2\,{\mu }^2} -
  \frac{i\,{\sigma }\,g \left( g -1 \right)
     }{{4\,\mu }^3} +
  \frac{g \left( g -1 \right) }{4\,{\mu }^4} +
  {{O}({\mu^{-5} })} }\, ,
\end{array}
\end{equation}
where $\sigma = 1$ for $\Im(\mu)>0$, and $\sigma = -1$ for
$\Im(\mu)<0$. The coefficients in this series can be
straightforwardly evaluated from Eqs.\ (\ref{P+Q}) and (\ref{T}).
Notice that $A_k(g,-1)=A_k(g,1)^*$, since $A_{2k}(g,1)$ is real
and $A_{2k+1}(g,1)$ is pure imaginary.

Similarly, from (\ref{JprimasobreJasymp}) we  simply get for the
second factor in the second term in the right hand side of Eq.\
(\ref{TrG2})
\begin{equation}\label{trdifasymp}
   Tr\{G_D(\mu^2) -
    G_N(\mu^2)\} \sim \frac{2 g -1}{2\, \mu^2} \, .
\end{equation}

\bigskip

Finally, taking into account Eq.\ (\ref{JsobreJup}), we have
\begin{equation}\label{tau-asymp}\begin{array}{c}
   \tau(\mu^2) \sim \displaystyle{\frac{1}
   {1 - \displaystyle{{e^
         {\sigma \,i\,
           \pi \left( g -\frac{1}{2}\right) }
           \,{\rho(\alpha,\beta)}\,{\mu }^{1-2\,g}}} }
        \sim }\\ \\  \sim
      \displaystyle{
    \sum_{k=0}^\infty \left(
    {e^{\sigma\, i\, \pi\, (g-\frac 1 2 )} \,
    \rho(\alpha,\beta)\,
    \mu^{1-2 g}}\right)^k \, ,}
\end{array}
\end{equation}
where $\sigma =1$ ($\sigma =-1$) corresponds to $\Im(\mu)>0$
($\Im(\mu)<0$).

Notice the appearance of $g$-dependent powers of $\mu$ in this
asymptotic expansion.

\bigskip

\section{The $\zeta$-function and the trace of the heat-kernel}
\label{spectral-functions}

The $\zeta$-function for a general self-adjoint extension of $D_x$
is defined, for $\Re(s)>1/2$, as
\begin{equation}\label{zeta}
  \zeta(s)=- \frac{1}{2\,\pi\,i} \oint_{\mathcal{C}}
  {\lambda^{-s}} \, Tr\left\{G(\lambda)\right\}
  \, d\lambda\, ,
\end{equation}
where the curve $\mathcal{C}$ encircles counterclockwise the
spectrum of the operator, keeping to the left of the origin.
According to Eq.\ (\ref{trazaG}), we have
\begin{equation}\label{zeta1}
  \zeta(s)= \zeta^D(s)+\frac{1}{2\,\pi\,i} \oint_{\mathcal{C}}
  {\lambda^{-s}} \,  \tau(\lambda)\,
  Tr\{G_D(\lambda)-G_N(\lambda)\}
  \, d\lambda\, ,
\end{equation}
where $\zeta^D(s)$ is the $\zeta$-function for the $D$-extension.

Since, according to the discussion in Section \ref{the-spectrum},
$D_x^D$ has a positive spectrum, and the self-adjoint extension
$D_x^{(\alpha,\beta)}$ has at most one negative eigenvalue, we can
write
\begin{equation}\label{zeta+}\begin{array}{c}
  \displaystyle{  \zeta^{(\alpha,\beta)}(s)=
  \zeta^{D}(s) +\Theta(s) \, -
    } \\ \\
    \displaystyle{
    -\frac{1}{2\,\pi\,i} \int_{-i\,\infty+0}^{i\,\infty+0}
  {\lambda^{-s}} \,
   \tau(\lambda)\,
  Tr\{G_D(\lambda)-G_N(\lambda)\}
  \, d\lambda
    } \, ,
\end{array}
\end{equation}
where $\Theta(s)=\lambda_-^{-s}$ if there is a negative
eigenvalue, and vanishes otherwise.

We can also write
\begin{equation}\label{zeta+1}\begin{array}{c}
  \displaystyle{\zeta^{(\alpha,\beta)}(s)=
    \frac{e^{-i\,\frac{\pi}{2}\,s}}{\pi}
     \int_{1}^{\infty}
  {\mu^{1-2s}} \,
  Tr\left\{G\Big((e^{i\,\frac{\pi}{4}}\, \mu)^2\Big) \right\}
  \, d\mu\,  +
  } \\ \\
  \displaystyle{
    +\frac{e^{ i\,\frac{\pi}{2}\,s}}{\pi} \int_{1}^{\infty}
  {\mu^{1-2s}} \,
  Tr\left\{G\Big((e^{-i\,\frac{\pi}{4}}\, \mu)^2\Big)\right\}
  \, d\mu + {h_1(s)} \, ,
  }
\end{array}
\end{equation}
where $h_1(s)$ is an entire function. Therefore, in order to
determine the poles of $\zeta^{(\alpha,\beta)}(s)$, we can
subtract and add a partial sum of the asymptotic expansion
obtained in the previous Section to $Tr\left\{G(\lambda)\right\}$
in the integrands in the right hand side of Eq.\ (\ref{zeta+1}).

\bigskip

In so doing, we get for the $D$-extension and for a real $s>1/2$
\begin{equation}\label{zetaD+1}\begin{array}{c}
  \displaystyle{ \zeta^{D}(s) =} \\ \\
  \displaystyle{ =
    \frac{1}{\pi}\sum_{\sigma=\pm 1} \int_{1}^{\infty}
    e^{-i\,\sigma\,\frac{\pi}{2}\,s}\,
  {\mu^{1-2 s}} \,
  \left\{ \sum_{k=1}^{N} e^{-i\,\sigma\,\frac{\pi}{4}\,k}\,
  A_k(g,\sigma) \, \mu^{-k} \right\}
  \, d\mu\,  +
  } \\ \\
  \displaystyle{ +{h_2(s)}
  = \frac{1}{\pi}\,\sum_{k=1}^{N}\frac{
  \Re \left\{
  e^{-i\,\frac{\pi}{2}\,(s+k/2)}\,A_k(g,1) \right\}
  }{s-(1-k/2)}
  + {h_2(s)}\, ,
  }
\end{array}
\end{equation}
where $h_2(s)$ is holomorphic in the open half plane $\Re(s)>
(1-N)/2$.

Consequently, the meromorphic extension of $\zeta^{D}(s)$ presents
simple poles at
\begin{equation}\label{poles-D}
    s=1-k/2\, ,\quad {\rm for}\  k=1,2,3,\dots,
\end{equation}
with residues
\begin{equation}\label{otros-residuos}
  \left. {\rm Res}\,\zeta^{D}(s) \right|_{s=1-k/2}
  =- \frac{1}{\pi}\,\Re\left\{i\, A_k(g,1)\right\}  \, ,
\end{equation}
where the coefficients $A_k(g,1)$ are given in Eq.\
(\ref{asymp-trGD-upp}). Notice that these residues vanish for even
$k$.

In particular, for $s=1/2$ ($k=1$) one gets
\begin{equation}\label{otros-residuos-s12}
  \left. {\rm Res}\,\zeta^{D}(s) \right|_{s=1/2}
  =- \frac{1}{\pi}\,\Re\left\{i\, A_1(g,1)\right\}=
  \frac{1}{2\, \pi}  \, .
\end{equation}
This is the unique pole present in $\zeta^{D}(s)$ for the $g=1$
case, where there is no singularity in the 0-th order coefficient
of $D_x$.

\bigskip

For a general self-adjoint extension $D_x^{(\alpha,\beta)}$, we
must also consider  the singularities coming from the asymptotic
expansion of $\tau(\lambda)\,Tr\{G_D(\lambda)-G_N(\lambda)\}$ in
Eq.\ (\ref{trazaG}), given in Eqs.\ (\ref{trdifasymp}) and
(\ref{tau-asymp}).

\bigskip

From Eq.\ (\ref{zeta+}), and taking into account Eq.\
(\ref{zeta+1}), for real $s>1/2$ we can write
\begin{equation}\label{zetadif}\begin{array}{c}
    \zeta^{(\alpha,\beta)}(s)-\zeta^{D}(s)=
    \displaystyle{{h_3(s)}} \, - \frac{ 2\,g-1}{2\,\pi}\, \times\\ \\
  \displaystyle{
     \sum_{\sigma=\pm 1}
          e^{-i\,\sigma\,\frac{\pi}{2}\,(s+1)}\,
 \int_{1}^{\infty}
  {\mu^{-1-2s}} \,
  \left\{ \sum_{k=0}^{N}\left({ e^{i\,\sigma\,
  \frac{\pi}{2}(g-\frac 1 2)}}
  {\rho(\alpha,\beta)}\,
   \mu^{1-2\,g} \right)^k\right\}
  \, d\mu \, =
  } \\ \\
  \displaystyle{
  -\left( \frac{2\,g-1}{2\,\pi}\right) \sum_{k=0}^{N}\,
  \frac{1}{s-(\frac 1 2 -g)k}\
  \Re \left\{
  {e^{i\,\frac{\pi}{2} \left((g-\frac 1 2)k-s-1\right)}}
  {\rho(\alpha,\beta)^k}\right\}
  + \displaystyle{{h_3(s)}}\, ,
  }
\end{array}
\end{equation}
where $h_3(s)$ is holomorphic for $\Re(s)>\left(\frac 1 2 -
g\right) (N+1)$.

Therefore, $\left(\zeta^{(\alpha,\beta)}(s)-\zeta^{D}(s)\right)$
has a meromorphic extension which presents  simple poles located
at negative $g$-dependent positions,
\begin{equation}\label{polos-zeta}
    s=-\left(g- \frac 1 2 \right) k\, ,
     \quad {\rm for}\  k=1,2,\dots\, ,
\end{equation}
with residues which depend on the self-adjoint extension given by
\begin{equation}\label{res-g-dep}\begin{array}{c}
      \left. {\rm Res}\,\left\{\zeta^{(\alpha,\beta)}(s)-
  \zeta^{D}(s)\right\} \right|_{s=\left(\frac 1 2 - g\right)k} =
    \\ \\
    = \displaystyle{-\left( \frac{2\,g-1}{2\, \pi}\right)
    \rho(\alpha,\beta)^k
    \ {\sin\left[\frac{\pi}{2}\left( 2g-1\right)k \right]}
  }\, .
\end{array}
\end{equation}

Notice that these poles are irrational for irrational values of
$g$. Moreover, the residues vanish for the ``N-extension" ($
\rho(\alpha,0)=0$), and have a singular limit for $\alpha
\rightarrow 0$.

\medskip

In particular, these poles for the $g=1$ case (for which there are
no singularity in the zero-th order term of $D_x$) are negative
half-integers, since in this case the residues vanish for even
$k$.

\medskip

It is interesting to notice that the poles in Eq.\
(\ref{polos-zeta}) are also poles of the $\zeta$-function of the
corresponding self-adjoint extension of the operator
$-\partial_x^2+{g(g-1)}\,{x^{-2}}+x^2$ in
$\mathbf{L}_2(\mathbb{R}^+)$ considered in \cite{FPW}, with
exactly the same residues, as can be easily verified.

\bigskip


{ Let us remark that when $\alpha \neq 0$ the residue of
$\zeta^{(\alpha,\beta)}$ at $s=-\left(g- \frac 1 2 \right) k$ is a
constant times $(\beta/\alpha)^{k}$. This is consistent with the
behavior of $D_x$ under the scaling isometry
$Tu(x)=c^{1/2}\,u(c\,x)$ taking $\mathbf{L_2}(0,1)\rightarrow
\mathbf{L_2}(0,1/c)$. The extension $D_x^{(\alpha,\beta)}$ is
unitarily equivalent to the operator
$(1/c^2)\dot{D}_x^{(\alpha',\beta')}$ similarly defined on
$\mathbf{L_2}(0,1/c)$, with $\alpha' = c^{-g}\, \alpha$ and
$\beta' = c^{g-1} \, \beta$:
\begin{equation}\label{isometry}
  T\,D_x^{(\alpha,\beta)} = \frac 1 {c^2} \,
  \dot{D}_x^{(\alpha',\beta')}\, T\, .
\end{equation}
Notice that only for the extensions with $\alpha=0$ or $\beta=0$
the boundary condition at the singular point $x=0$, Eq.\
(\ref{BC2}), is left invariant by this scaling.

Therefore, we have for the  $\zeta$-function of the scaled problem
\begin{equation}\label{zetas-isometry}
  \dot{\zeta}^{(\alpha',\beta')}(s)=
   c^{-2\, s}\,\zeta^{(\alpha,\beta)}(s)\, ,
\end{equation}
and for the residues
\begin{equation}\label{zetas-isometry-residues}
   \left. {\rm Res}\,\left\{\dot{\zeta}^{(\alpha',\beta')}(s)
   \right\} \right|_{s=\left(\frac 1 2 -g\right)k} = c^{(2\,g -1)k}
   \left. {\rm Res}\,\left\{{\zeta}^{(\alpha,\beta)}(s)
   \right\} \right|_{s=\left(\frac 1 2 -g\right)k}\, .
\end{equation}
The factor $c^{(2\,g -1)k}$ exactly cancels the effect the change
in the boundary condition at the singularity has on
$\rho(\alpha,\beta)$,
\begin{equation}\label{change-in-rho}
  \rho(\alpha,\beta)^{k}
  =c^{(1-2\,g)k}\, \rho(\alpha',\beta')^{k}\, .
\end{equation}
Then, the difference between the intervals $(0,1)$ and $(0,1/c)$
has no effect on the structure of these residues, which presumably
are determined locally in a neighborhood of $x=0$. }

\bigskip


In this way we conclude that, for a general self-adjoint
extension, the presence  of poles in the $\zeta$-function located
at $g$-dependent positions is a consequence of the singular
behavior ($\sim x^{-2}$) of the zero-th order term in $D_x$ near
the origin, together with a scaling non-invariant boundary
condition at the singularity.

\bigskip

Finally, let us remark that the relation between the
$\zeta$-function and the trace of the heat-kernel of
$D_x^{(\alpha,\beta)}$,
\begin{equation}\label{heat}
  \zeta^{(\alpha,\beta)}(s)=\frac{1}{\Gamma(s)}\int_0^1
  t^{s-1}\, Tr\left\{e^{-t\,D_x^{(\alpha,\beta)}}\right\}\, dt
  + H(s)\, ,
\end{equation}
where $H(s)$ is an entire function, straightforwardly lead to the
following small-$t$ asymptotic expansion,
\begin{equation}\label{heat-asymp}\begin{array}{c}
  \displaystyle{  Tr\left\{e^{-t\,D_x^{(\alpha,\beta)}}-
  e^{-t\,D_x^{D}}\right\}
  \sim  \left(g-\frac 1 2\right) -} \\ \\
  \displaystyle{- \sum_{k=1}^\infty}
  \left\{\Gamma\left(\left[\frac 1 2-g \right]k\right)
  \frac{2\,g-1}{2\, \pi}\,\rho(\alpha,\beta)^k
    \ {\sin\left[
    \frac{\pi}{2}\left( 2g-1\right)k \right]}\right\}
    t^{\left(g-\frac{1}{2}\right)k}\, .
\end{array}
\end{equation}
The first term in the right hand side, coming from Eq.\
(\ref{trdifasymp}) and the first term in the asymptotic expansion
of $\tau(\lambda)$ in Eq.\ (\ref{tau-asymp}), coincides with the
result reported in \cite{Mooers}. Notice also the $g$-dependent
powers of $t$ appearing in the asymptotic series in the right hand
side of Eq.\ (\ref{heat-asymp}) for any general self-adjoint
extension (except for the ``N-extension", for which
$\rho(\alpha,0)=0$). In particular, the first term in this series
reduces to
\begin{equation}\label{first-term}
    - \frac{\beta}{\alpha}\, \frac{2^{2g-1}}{\Gamma(\frac 1 2 -g)}\
     t^{g-\frac 1 2}\, .
\end{equation}
This power of $t$ also coincides with the result quoted in
\cite{Mooers}, but we find a different coefficient.

\bigskip

\noindent {\bf Acknowledgements:} We would like to thank Prof.\
Robert Seeley for useful discussions.

\medskip

H.F.\ and P.A.G.P.\ acknowledge support from Universidad Nacional
de La Plata (grant 11/X298) and CO\-NI\-CET (grant 0459/98),
Argentina.

M.A.M.\ acknowledge support from Universidad Nacional de La Plata
(grant 11/X228), Argentina.

\appendix


\bigskip

\section{The case $g=1/2$ \label{g=1/2}}

The case $g=1/2$, for which the differential operator $D_x$ takes
the form
\begin{equation}\label{g1/2}
    D_x=-\frac{d^2}{dx^2}-\frac{1}{4\, x^2}\, ,
\end{equation}
requires a separate consideration which we briefly present in this
Appendix.

\bigskip

Along the same lines as in the proof of Lemma \ref{lema1-1}, it is
straightforward to show that, if $\phi(x)\in \mathcal{D}(D_x^*)$,
then
\begin{equation}\label{l1C-1}
    \displaystyle{\left|\,\phi(x)-\left(C_1[\phi]\, \sqrt{x} +
  C_2[\phi]\, \sqrt{x}\, \log{x}\right)\right|
  \leq \frac{\|D_x\phi(x)\|}{\sqrt{2}} \  x^{3/2}}
\end{equation}
and
\begin{equation}\label{l1C-2}\begin{array}{c}
  \displaystyle{
      \left|\,\phi'(x)-\left[\frac{1}{2}C_1[\phi]\,x^{-1/2}+
      C_2[\phi]\left(x^{-1/2}+\frac{1}{2}x^{-1/2}\log{x}
      \right)\right]\right| \leq} \\ \\
  \displaystyle{\leq
  \frac{3}{2\sqrt{2}}\|D_x\phi(x)\|\, x^{1/2}}
\end{array}
\end{equation}
for some constants $C_1[\phi]$ and $C_2[\phi]$, where $\| \cdot
\|$ stands for the $\mathbf{L_2}$-norm.

\medskip

Therefore, it is easy to see that Eq.\ (\ref{DDstar}) is also
valid in the present case, and the self-adjoint extensions of
$D_x$ correspond again to those subspaces $S\subset \mathbb{C}^4$
such that $S=S^{\perp}$, with the orthogonal complement taken in
the sense of the symplectic  form on the right hand side of Eq.\
(\ref{DDstar}).

\medskip

If, in addition, we select the Dirichlet condition at $x=1$,
$\phi(1)=0$, the remaining self-adjoint extensions of $D_x$
correspond to a one-parameter family characterized by Eq.\
(\ref{BC2}), $D_x^{(\alpha,\beta)}$.

\medskip

There exists a particular self-adjoint extension for which
$C_2[\phi]=0$, namely $D_x^D:=D_x^{(0,1)}$, such that  the
functions in its domain behave near the origin as
\begin{equation}
    \phi(x)=C_1[\phi]\,\sqrt{x}+O(x^{3/2}).
\end{equation}
The eigenfunction of $D_x^D$ corresponding to the eigenvalue
$\lambda$ is given by,
\begin{equation}
    \phi(x)=C_1[\phi]\sqrt{x}\, J_0(\mu x),
\end{equation}
where $\lambda=\mu^2$ and $\mu$ is a (positive) zero of
$J_0(\mu)$.

\medskip

For an arbitrary self-adjoint extension $D_x^{(\alpha,\beta)}$
with $\alpha\neq 0$, the eigenfunction corresponding to the
eigenvalue $\lambda=\mu^2$ is given by
\begin{equation}\begin{array}{c}
  \phi(x)=\left\{C_1[\phi]-C_2[\phi](\log{\mu}-\log{2}+\gamma)
    \right\}\,\sqrt{x}\,J_0(\mu x)\, + \\ \\ \displaystyle{ +
  \frac{\pi}{2}\, C_2[\phi]\,\sqrt{x}\,N_0(\mu x) }\, ,
\end{array}
\end{equation}
where $C_1[\phi],C_2[\phi]$ are constrained by eq.\ (\ref{BC2}).
The condition $\phi(1)=0$ leads to the equation
\begin{equation}
    (\theta-\log{\mu})J_0(\mu)+
    \frac{\pi}{2}N_0(\mu)=0\, ,
\end{equation}
where $\theta=-\beta/\alpha+\log{2}-\gamma$, which determines the
spectrum of $D_x^{(\alpha,\beta)}$. Notice that there are no
negative eigenvalues.

\bigskip

In order to determine the kernels of the resolvents
$\mathcal{G}^D(\mu^2):=(D_x^{D}-\mu^2)^{-1}$ and
$\mathcal{G}^{(\alpha,\beta)}(\mu^2):=(D_x^{(\alpha,\beta)}-\mu^2)^{-1}$,
we define
\begin{equation}\label{LyR} \left\{
\begin{array}{l}
  \mathcal{L}^{D}(x;\mu)=\sqrt{x}\, J_0(\mu x) \, ,\\ \\
  \displaystyle{
  \mathcal{L}^{(\alpha,\beta)}(x;\mu)=\sqrt{x}\,
  \left\{(\theta-\log{\mu})J_0(\mu x)+
    \frac{\pi}{2}N_0(\mu x)\right\}\, ,}\\ \\
    \mathcal{R}(x;\mu)=\sqrt{x}\, \left\{N_0(\mu)\,J_0(\mu
    x)-J_0(\mu)\,N_0(\mu x)\right\}\, ,
\end{array} \right.
\end{equation}
to get
\begin{equation}\begin{array}{c}
  \mathcal{G}^{D}(x,y;\mu^2)= \\ \\
  \displaystyle{
  =\frac{1}
    {W[\mathcal{L}^{D}(x;\mu),\mathcal{R}(x;\mu)]}
    \left\{\begin{array}{lr}
    \mathcal{L}^{D}(x;\mu)\,\mathcal{R}(y;\mu)\, , & x\leq y \, ,
    \\ \\
    \mathcal{L}^{D}(y;\mu)\,\mathcal{R}(x;\mu)\, , & x\geq y \, ,
    \end{array}\right.}
\end{array}
\end{equation}
and
\begin{equation}\begin{array}{c}
  \mathcal{G}^{(\alpha,\beta)}(x,y;\mu^2)= \\ \\
  \displaystyle{
  =\frac{1}
    {W[\mathcal{L}^{(\alpha,\beta)}(x;\mu),\mathcal{R}(x;\mu)]}
    \left\{\begin{array}{lr}
    \mathcal{L}^{(\alpha,\beta)}(x;\mu)\,\mathcal{R}(y;\mu)\, , & x\leq y\, ,
    \\ \\
    \mathcal{L}^{(\alpha,\beta)}(y;\mu)\,\mathcal{R}(x;\mu)\, , & x\geq y\,
    ,
    \end{array}\right.}
\end{array}
\end{equation}
where the Wronskians can be easily computed from (\ref{LyR}),
\begin{equation}\begin{array}{c}
    \displaystyle{
  W[\mathcal{L}^{D}(x;\mu),\mathcal{R}(x;\mu)]=
    \frac{2}{\pi}\, J_0(\mu) }\, ,\\ \\
    \displaystyle{
  W[\mathcal{L}^{(\alpha,\beta)}(x;\mu),
    \mathcal{R}(x;\mu)]=\frac{2}{\pi}\,
    (\theta-\log{\mu})J_0(\mu)+
    N_0(\mu)}\, .
\end{array}
\end{equation}

From Eq.\ (\ref{large-zeroes}), it can be seen that both
$\mathcal{G}^D(\lambda)$ and
$\mathcal{G}^{(\alpha,\beta)}(\lambda)$ are trace class operators.

\bigskip

Now, taking into account that \cite{A-S,Mathematica}
\begin{equation}\label{primi-0}\begin{array}{c}
  \displaystyle{\int x\, Z_1(0,x)\, Z_2(0, x)\, dx =} \\ \\
  \displaystyle{=\frac{x^2 }{2} \left\{Z_1(0,x)\, Z_2(0, x)
    +Z_1(1,x)\, Z_2(1, x)
    \right\}\, ,  }
\end{array}
\end{equation}
where $Z_{1,2}(\nu,x)= J_\nu(x)$ or $N_\nu(x)$, the traces of the
resolvents can be readily computed to get
\begin{equation}\label{traces-0}\begin{array}{c}
    \displaystyle{
  Tr\left\{\mathcal{G}^{D}(\mu^2)\right\}=\int_0^1
    \mathcal{G}^{D}(x,x;\mu^2)\,dx
    =\frac{1}{2\mu}\frac{J_1(\mu)}{J_0(\mu)} }\, ,\\ \\
    \displaystyle{
  Tr\left(\mathcal{G}^{(\alpha,\beta)(\mu^2)}\right)=\int_0^1
    \mathcal{G}^{(\alpha,\beta)}(x,x;\mu^2)\,dx=  }\\ \\
    \displaystyle{
    = \frac{1}{2\mu}\, \frac{\frac{2}{\pi}(\theta-\log{\mu})J_1(\mu)+N_1(\mu)}
    {\frac{2}{\pi}(\theta-\log{\mu})J_0(\mu)+N_0(\mu)} }\, .
\end{array}
\end{equation}

\medskip

From Eqs.\ (\ref{Jupper} - \ref{N-sigma}) one straightforwardly
gets the same asymptotic expansion for these two traces,
\begin{equation}\label{tr-gd-0}
    \begin{array}{c}
    \displaystyle{
      Tr\left\{\mathcal{G}^{D}(\mu^2)\right\}\sim
    \frac{e^{i\sigma \frac \pi 2}}{2\mu} \left(
    \frac{P(1,\mu)- i \sigma \, Q(1,\mu)}
    {P(0,\mu)- i \sigma \, Q(0,\mu)}\right)
    \sim
    Tr\left(\mathcal{G}^{(\alpha,\beta)}(\mu^2)\right) \sim }\\ \\
    \displaystyle{\sim \sum_{k=1}^\infty \frac{A_k(1/2,\sigma)}{\mu^k}
      = {\frac{i\,\sigma }{2\mu } } +
  \frac{1}{4\,{\mu }^2} +
  {\frac{i \,\sigma}{16{\mu }^3} } -
  \frac{1}{16\,{\mu }^4} +
  {{O}({\mu^{-5} })} }\, ,
    \end{array}
\end{equation}
where $\sigma = +1$ ($-1$) for $\Im(\mu)>0$ ($\Im(\mu)<0$).

Notice that the asymptotic series in Eq.\ (\ref{tr-gd-0})
coincides with the right hand side of Eq.\ (\ref{asymp-trGD-upp})
evaluated at $g=1/2$. Therefore, from Eq.\ (\ref{zetaD+1}) one
concludes that, in the present case, $\zeta^{(\alpha,\beta)}(s)$
has simple poles only at $s=1-k/2$, for $k=1,2,3,\dots$, with
residues given by
\begin{equation}\label{residuos-D-0}
  \left. {\rm Res}\,\zeta^{D}(s) \right|_{s=1-k/2}
  =- \frac{1}{\pi}\,\Re\left\{i\, A_k(1/2,1)\right\}
\end{equation}
(vanishing for even $k$) for all the self-adjoint extensions of
$D_x$.

\medskip

So, in contrast to the case of $1/2<g<3/2$, the pole structure of
the $\zeta$-function for $g=1/2$ is independent of the
self-adjoint extension considered and does not differ from the
usual one.


\section{Evaluation of the traces of the resolvents}\label{integrals}

In this Appendix we briefly describe the evaluation of the traces
appearing in Section \ref{trace-resolvent}.

From Eq.\ (\ref{GD11}) we get for the kernel of $G_D(\lambda)$ on
the diagonal
\begin{equation}\label{trGDDD}\begin{array}{c}
    G_D(x,x;\mu^2)
  =    \gamma_D\, x \left\{ J_{\frac{1}{2} - g}(\mu )\,
 J_{g-\frac{1}{2}}(\mu \, x)^2 - \right.\\ \\
 \left. J_{g-\frac{1}{2}}(\mu ) \,
 J_{g-\frac{1}{2}}(\mu\, x ) \, J_{\frac{1}{2} - g}(\mu\, )\right\}
 \, .
\end{array}
\end{equation}
Therefore, in order to evaluate its trace it is sufficient to know
the primitives \cite{Mathematica,GR}
\begin{equation}\label{nu-nu}
  \int x\, J_\nu^2(\mu \,x)\,dx
  =\frac{x^2}{2}\,\left\{ {J_\nu(x\,\mu  )}^2 -
      J_{\nu-1}(x\,\mu  )\,
       J_{\nu+1}(x\,\mu  ) \right\}
\end{equation}
and
\begin{equation}\label{nu-menosnu}\begin{array}{c}
 \displaystyle{ \int x\, J_\nu(\mu \,x)\,
 J_{-\nu}(\mu \,x)\,dx= }\\ \\
     =\displaystyle{\frac{ - {\nu }^2}
      {{\mu  }^2\,
      \Gamma(1 - \nu )\,
      \Gamma(1 + \nu )}\,
      \left[ \,_1 F_2
          \left(\{ - {1}/
            {2}  \} ,
         \{ -\nu ,\nu \} ,
         -x^2 \, {\mu  }^2\right)-1  \right]}\, ,
\end{array}
\end{equation}
where
\begin{equation}\label{1F2}\begin{array}{c}
   _1 F_2 \left(\{ - {1}/
            {2}  \} ,
         \{ -\nu ,\nu \} ,
         -x^2 \, {\mu  }^2\right)
  \displaystyle{
  =- \frac{ \pi \,x^2\,
      {\mu  }^2\,
       \,\csc (\pi \,\nu )
       }{4\,\nu }\, \times }\\ \\
      \left\{ J_{-1 - \nu }(x\,\mu  )\,
         J_{-1 + \nu }(x\,\mu  ) +
        2\,J_{-\nu }(x\,\mu  )\,
         J_{\nu }(x\,\mu  ) +
        J_{1 - \nu}(x\,\mu  )\,
         J_{1 + \nu }(x\,\mu  )
        \right\}\, .
\end{array}
\end{equation}

These primitives, together with the relation
\begin{equation}\label{prop-J-Bessel}
  J_{\nu-1}(z)+J_{\nu+1}(z)
  =\frac{2 \nu}{z} J_{\nu}(z)\, ,
\end{equation}
necessary to simplify the intermediate results, straightforwardly
lead to Eq.\ (\ref{traza-GD}).

\bigskip

Similarly, for the kernel of $G_N(\lambda)$ on the diagonal we
have
\begin{equation}\label{trGD}\begin{array}{c}
     G_N(x,x;\mu^2) = \gamma_N\, x
     \left\{
     -J_{g-\frac{1}{2}}(\mu ) \,
     {J_{\frac{1}{2} - g}(\mu\,x )}^2\,
      +  \right. \\ \\ \left.
     +   J_{\frac{1}{2} - g}(\mu )\,
   J_{\frac{1}{2} - g}(\mu\, x )\,
   J_{g-\frac{1}{2}}(\mu \,x) \right\}
    \, .
\end{array}
\end{equation}
The same argument as before leads to Eq.\ (\ref{traza-GN}).

\bigskip

\section{The Hankel expansion}\label{Hankel}

To develop an asymptotic expansion for the trace of the resolvent
we  employ  the Hankel asymptotic expansion for the Bessel
functions which, for completeness, we briefly describe in this
Appendix.

\medskip

For $|z|\rightarrow \infty$, with $\nu$ fixed and $|\arg z|<\pi$,
we have \cite{A-S}
\begin{equation}\label{hankel}
  J_\nu(z)\sim \left(\frac{2}{\pi\,z}\right)^{\frac 1 2}
  \left\{ P(\nu,z) \cos \chi(\nu,z)
  -Q(\nu,z) \sin \chi(\nu,z) \right\}\, ,
\end{equation}
and
\begin{equation}\label{hankel-N}
    N_\nu(z)\sim \left(\frac{2}{\pi\,z}\right)^{\frac 1 2}
  \left\{ P(\nu,z) \sin \chi(\nu,z)
  + Q(\nu,z) \cos \chi(\nu,z) \right\}\, ,
\end{equation}
where
\begin{equation}\label{chi}
  \chi(\nu,z) = z- \left(\frac \nu 2 + \frac 1 4\right) \pi,
\end{equation}
\begin{equation}\label{P}
  P(\nu,z) \sim \sum_{k=0}^\infty
  \frac{(-1)^k\,\Gamma\left(
  \frac 1 2 + \nu + 2 k\right)}{(2k)! \, \Gamma\left(
  \frac 1 2 + \nu - 2 k\right)} \,
  \frac{1}{\left(2 z\right)^{2 k}}\, ,
\end{equation}
and
\begin{equation}\label{Q}
  Q(\nu,z) \sim \sum_{k=0}^\infty
  \frac{(-1)^k \,\Gamma\left(\frac 1 2+
   \nu + 2 k+1 \right)}{(2 k+1)! \, \Gamma\left(
   \frac 1 2+ \nu - 2 k-1 \right)} \,
    \frac{1}{\left(2 z\right)^{2 k+1}}\, .
\end{equation}

Moreover, $P(-\nu,z)=P(\nu,z)$ and $Q(-\nu,z)=Q(\nu,z)$, since
these functions depend only on $\nu^2$ (see Ref.\ \cite{A-S}, page
364).

Therefore,
\begin{equation}\label{Jupper}
  J_\nu(z)\sim
  \frac{e^{-i\sigma z}\, e^{i\sigma \pi \left(
  \frac \nu 2 + \frac 1 4 \right)}}{\sqrt{2\pi z}}
  \left\{ P(\nu,z)
  - i\sigma \,Q(\nu,z) \right\}\, ,
\end{equation}
where $\sigma=1$ for $z$ in the upper open  half plane and $\sigma
=-1$ for $z$ in the lower open  half plane.

Similarly,
\begin{equation}\label{N-sigma}
    N_\nu(z)\sim i \sigma\, \frac{e^{-i \sigma z}
    \, e^{i \sigma  \pi \left(
  \frac \nu 2 + \frac 1 4 \right)}}{\sqrt{2 \pi z}}\,
  \left\{ P(\nu,z)
  - i \sigma \,Q(\nu,z) \right\}\, ,
\end{equation}
with $\sigma =1$ if $\Im(z)>0$ and $\sigma=-1$ for $\Im(z)<0$.

In these equations,
\begin{equation}\label{P+Q}
  P(\nu,z) - i\sigma  \,Q(\nu,z)\sim
  \sum_{k=0}^\infty \langle \nu , k\rangle
  \, \left(\frac{- i \sigma}{2 z}\right)^k\, ,
\end{equation}
where the coefficients
\begin{equation}\label{coef-hankel}
  \langle \nu , k\rangle=
  \frac{\Gamma\left(\frac 1 2 +\nu + k\right)}
  {k! \, \Gamma\left(\frac 1 2 +\nu - k\right)}
  =\langle -\nu , k\rangle
\end{equation}
are the Hankel symbols.

\bigskip

For the quotient of two Bessel functions we have
\begin{equation}\label{JsobreJup1}
      \frac{J_{\nu_1}(z)}{J_{\nu_2}(z)}
      \sim e^{ i\sigma \frac{\pi}{2} (\nu_1 - \nu_2)}\,
      \frac{P(\nu_1 ,z)- i\sigma \,
      Q(\nu_1 ,z) }
      {P(\nu_2,z) - i\sigma \,
      Q(\nu_2,z) }\, ,
\end{equation}
where $\sigma =1$ for $\Im(z)>0$ and $\sigma=-1$ for $\Im(z)<0$.
The coefficients of this asymptotic expansion can be easily
obtained, to any order, from Eq.\ (\ref{P+Q}),
\begin{equation}\label{asymp-cociente}\begin{array}{c}
    \displaystyle{\frac{P(\nu_1 ,z)\pm i\,
      Q(\nu_1 ,z) }
      {P(\nu_2,z) \pm i\,
      Q(\nu_2,z) }} \sim
      1 + \Big( \langle \nu_1 , 1\rangle -
     \langle \nu_2 , 1\rangle \Big) \,\left(\frac{\pm i}{2 z}\right)
      +  O\left(\frac 1 {z^2}\right)\, .
\end{array}
\end{equation}

In particular,
 \begin{equation}\label{JsobreJup}
      \frac{J_{\frac 1 2 -g}(z)}{J_{g-\frac 1 2}(z)}
      \sim e^{ i\sigma \pi \left(\frac 1 2 -g\right)}
      \frac{P(\frac 1 2 -g,z)- i \sigma \,
      Q(\frac 1 2 -g,z) }
      {P(g-\frac 1 2,z) - i\sigma \,
      Q(g-\frac 1 2,z) } = e^{ i \sigma  \pi \left(\frac 1 2
      -g\right)}\, ,
\end{equation}
since $P(\nu,z)$ and $Q(\nu,z)$ are even in $\nu$.

\bigskip

Similarly, the derivative of the Bessel function has the following
asymptotic expansion \cite{A-S} for $|\arg z|<\pi$,
\begin{equation}\label{deriv-asymp}
  J'_\nu(z) \sim -\frac{2}{\sqrt{2 \pi  z}}
  \left\{
  R(\nu,z) \sin \chi(\nu,z) + S(\nu,z) \cos\chi(\nu,z)
  \right\}\, ,
\end{equation}
and
\begin{equation}\label{deriv-asymp-N}
  N'_\nu(z) \sim \frac{2}{\sqrt{2 \pi  z}}
  \left\{
  R(\nu,z) \cos \chi(\nu,z) - S(\nu,z) \sin \chi(\nu,z)
  \right\}\, ,
\end{equation}
where
\begin{equation}\label{R}
  R(\nu,z) \sim \sum_{k=0}^\infty(-1)^k\,
  \frac{\nu^2 + (2k)^2-1/4}
  {\nu^2-(2k-1/2)^2}\,
  \frac{\langle\nu,2k\rangle
  }{\left(2 z\right)^{2 k}}\, ,
\end{equation}
and
\begin{equation}\label{S}
  S(\nu,z) \sim \sum_{k=0}^\infty(-1)^k \,
  \frac{\nu^2 + (2k+1)^2-1/4}
  {\nu^2-(2k+1-1/2)^2}\,
  \frac{\langle\nu,2k+1\rangle
  }{\left(2 z\right)^{2 k+1}}\, .
\end{equation}
Then,
\begin{equation}\label{Jprima}
  J'_\nu(z)\sim \mp i\,
  \frac{e^{\mp i z}\, e^{\pm i \pi \left(
  \frac \nu 2 + \frac 1 4 \right)}}{\sqrt{2\pi z}}
  \left\{ R(\nu,z)
  \mp i \,S(\nu,z) \right\}\, ,
\end{equation}
where the upper sign is valid for $\Im(\lambda)>0$, and the lower
one for $\Im(\lambda)<0$. We have also
\begin{equation}\label{RS}
  R(\nu,z) \pm i \,S(\nu,z)=
  P(\nu,z) \pm i \,Q(\nu,z)
  + T_\pm(\nu,z)\, ,
\end{equation}
with
\begin{equation}\label{T}
  T_\pm(\nu,z)\sim
  \sum_{k=1}^\infty
  (2k-1)\langle\nu,k-1\rangle
  \left(\frac{\pm i}{2z}\right)^{k}\, .
\end{equation}

Therefore, we get
\begin{equation}\label{JprimasobreJ}
    \frac{J'_\nu(z)}{J_\nu(z)}
  \sim \mp  i \left\{
  1+ \frac{T_\mp(\nu,z)}{P(\nu,z)\mp i Q(\nu,z)}\right\}\, ,
\end{equation}
where the upper sign is valid for $\Im(\lambda)>0$, and the lower
one for $\Im(\lambda)<0$. The coefficients of the asymptotic
expansion in the right hand side of Eq.\ (\ref{JprimasobreJ}) can
be easily obtained from Eq.\ (\ref{P+Q}) and (\ref{T}),
\begin{equation}\label{TsobrePQ}\begin{array}{c}
    \displaystyle{
   \frac{T_\pm(\nu,z)}{P(\nu,z)\pm i Q(\nu,z)}
  = \left(\frac{\pm i}{2 z}\right)
  + O\left(\frac 1 {z^2}\right)
  }
\end{array}
\end{equation}

Finally, since the Hankel symbols are even in $\nu$ (see Eq.\
(\ref{coef-hankel})), from Eq.\ (\ref{P+Q}), (\ref{T}) and
(\ref{JprimasobreJ}) we have
\begin{equation}\label{JprimasobreJasymp}
  \frac{J'_\nu(z)}{J_\nu(z)}
  \sim \frac{J'_{-\nu}(z)}{J_{-\nu}(z)}\, .
\end{equation}


\end{document}